\documentclass[aps,prl,reprint,superscriptaddress]{revtex4-1}

\usepackage{graphicx} 
\usepackage{color}

\begin{document}


\title{Hourglass dispersion and resonance of magnetic excitations in the superconducting state of the single-layer cuprate HgBa$_2$CuO$_{4+\delta}$ near optimal doping}


\author{M. K. Chan}
\email{mkchan@lanl.gov}
\affiliation{School of Physics and Astronomy, University of Minnesota, Minneapolis, Minnesota 55455, USA}
\affiliation{Pulsed Field Facility, National High Magnetic Field Laboratory, Los Alamos National Laboratory, Los Alamos, New Mexico 87545, USA}

\author{Y. Tang}
\affiliation{School of Physics and Astronomy, University of Minnesota, Minneapolis, Minnesota 55455, USA}

\author{C. J. Dorow}
\altaffiliation{Present address: Department of Physics, University of California, San Diego, 9500 Gilman Drive La Jolla, CA 92093, USA}
\affiliation{School of Physics and Astronomy, University of Minnesota, Minneapolis, Minnesota 55455, USA}

\author{J. Jeong}
\affiliation{Laboratoire L{\'e}on Brillouin, CEA-CNRS, CEA-Saclay, 91191 Gif sur Yvette, France}

\author{L. Mangin-Thro}
\altaffiliation{Present address: Institute Laue Langevin, Grenoble 38042 CEDEX 9, France}
\affiliation{Laboratoire L{\'e}on Brillouin, CEA-CNRS, CEA-Saclay, 91191 Gif sur Yvette, France}

\author{M. J. Veit}
\altaffiliation{Present address: Department of Applied Physics, Stanford University, Stanford, CA 94305, USA }
\affiliation{School of Physics and Astronomy, University of Minnesota, Minneapolis, Minnesota 55455, USA}

\author{Y. Ge}
\altaffiliation{Present address: Department of Physics, Penn State University, University Park, PA 16802, USA}
\affiliation{School of Physics and Astronomy, University of Minnesota, Minneapolis, Minnesota 55455, USA}

\author{D. L. Abernathy}
\affiliation{Quantum Condensed Matter Division, Oak Ridge National Laboratory, Oak Ridge, Tennessee 37831, USA}

\author{Y. Sidis}
\affiliation{Laboratoire L{\'e}on Brillouin, CEA-CNRS, CEA-Saclay, 91191 Gif sur Yvette, France}

\author{P. Bourges}
\affiliation{Laboratoire L{\'e}on Brillouin, CEA-CNRS, CEA-Saclay, 91191 Gif sur Yvette, France}

\author{M. Greven}
\email{greven@umn.edu}
\affiliation{School of Physics and Astronomy, University of Minnesota, Minneapolis, Minnesota 55455, USA}

\date{\today}
\begin{abstract}
We use neutron scattering to study magnetic excitations near the antiferromagnetic wave vector in the underdoped single-layer cuprate HgBa$_2$CuO$_{4+\delta}$ (superconducting transition temperature $T_c \approx 88$~K, pseudogap temperature $T^* \approx 220$~K). The response is distinctly enhanced below $T^*$ and exhibits a Y-shaped dispersion in the pseudogap state, whereas the superconducting state features an X-shaped (hourglass) dispersion and a further resonance-like enhancement. A large spin gap of about 40 meV is observed in both states. This phenomenology is reminiscent of that exhibited by bilayer cuprates. The resonance spectral weight, irrespective of doping and compound, scales linearly with the putative binding energy of a spin-exciton described by an itinerant-spin formalism. 

\end{abstract}


\maketitle

The dynamic magnetic susceptibility of the hole-doped cuprates exhibits an hourglass-shaped (or X-shaped, upon considering an energy-momentum slice through ${\bf q}_{\rm AF}$) spectrum centered at the two-dimensional antiferromagnetic (AF) wave-vector ${\bf q}_{\rm AF}$~\cite{reznick04,tranquada04,pailhes04b}. Although the upper dispersive branch likely results from short-range AF correlations of local moments, the cause of the downward dispersive branch, at energies below the neck of the hourglass, has remained unclear. Results for the two cuprate families most widely studied via neutron scattering, (La,Nd)$_{2-x}$(Sr,Ba)$_x$CuO$_4$ (La214) and YBa$_2$Cu$_3$O$_{6+\delta}$ (Y123), support contradictory scenarios. For moderately- to overdoped Y123  (hole concentration $p\gtrsim0.085$), the low-energy  dispersion is accompanied by a magnetic resonance: an increase in scattering at ${\bf q}_{\rm AF}$ and energy $\omega_r$~\cite{mignod91,mook93}. Both features appear in the superconducting (SC) state and can be understood, within an {\it itinerant} picture, as a dispersive spin exciton bound below the particle-hole continuum and associated with the $d$-wave SC gap~\cite{norman01,eschrig06}. In contrast, La214 features an hourglass dispersion in both the SC and normal states, and no resonance in the SC state~\cite{tranquada04,lipscombe09,fujita12}. The discovery of static order of spatially segregated {\it localized} charge-spin stripes in La214~\cite{tranquada95} has motivated an interpretation in terms of fluctuating stripes~\cite{tranquada07}. Reconciliation of these discrepancies has been further complicated by the disparate crystal structures of Y123, a double-layer cuprate (two CuO$_2$ layers per primitive cell), and La214, a single-layer compound.

The subsequent observation of a magnetic resonance in single-layer Tl$_2$Ba$_2$Cu$_2$O$_{6+\delta}$ (Tl2201)~\cite{he02} and HgBa$_2$CuO$_{4+\delta}$ (Hg1201)~\cite{yu10}, which feature optimal $T_c$ values of nearly 100 K, more than twice that of La214, raised the prospect of a universal description of the magnetic response. However, detailed results have been difficult to obtain for these single-layer cuprates, and an hourglass dispersion has  not been detected. Thus, a connection, or lack thereof, between the hourglass dispersion, the resonance, and superconductivity has not been universally established, rendering a satisfactory description of magnetic excitations of single- and double-layer cuprates elusive.  

\begin{figure*}[t]
\includegraphics[width=.76\textwidth]{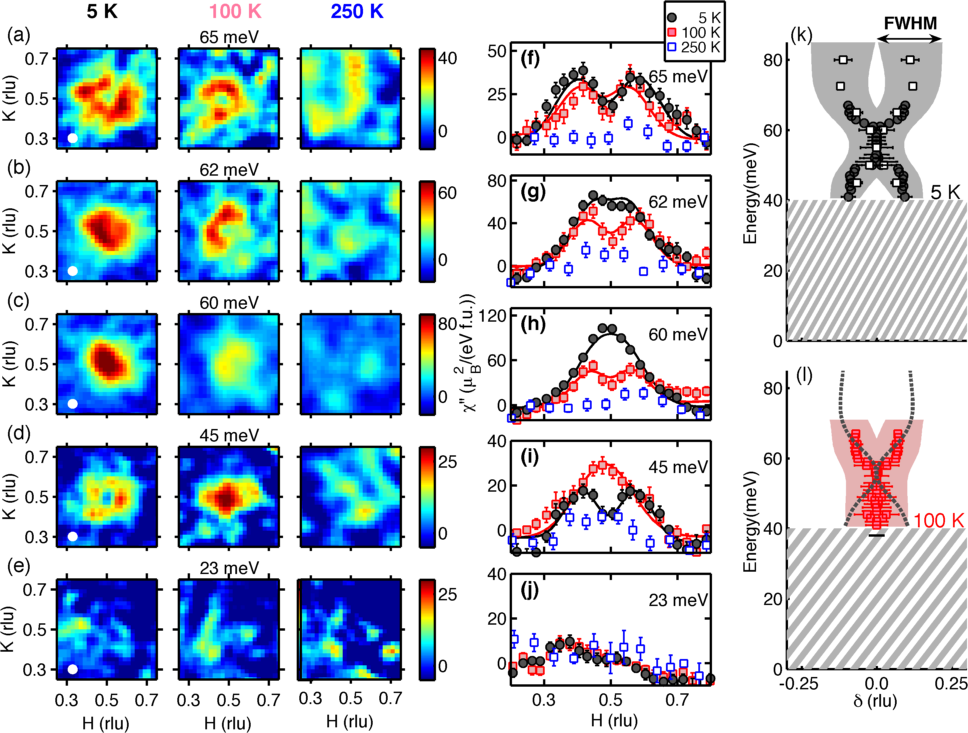}
\caption{
(a)-(e) Constant-energy images of magnetic susceptibility at $T = 5$~K (left), 100 K (middle), and 250 K (right). Data within a $6$~meV window centered at the indicated energies are averaged, except for $\omega = 23$~meV, where a 10 meV window was used. White dots (left-most panels): momentum resolution at each energy. (f)-(j) Corresponding constant-energy cuts averaged over $\{100\}$ and $\{010\}$ trajectories across ${\bf q}_{\rm AF}$.
Solid lines: gaussian fits to data convolved with the momentum resolution. (k) Energy dependence of incommensurability $\delta$ at $5$~K. Horizontal error bars: fit uncertainties for $\delta$. Filled black circles and open squares: data taken with incident energy $E_i = 100$~meV and  130 meV, respectively. Filled grey region: FWHM of the response. Hatched area: magnetic excitation gap. (l) Energy dependence of incommensurability $\delta$ at $100$~K, with dispersion at $5$~K (dotted line) shown for comparison. Horizontal black bar: experimental momentum resolution at $\omega = 40$~meV. \label{slice}}
\end{figure*}

A recent study of underdoped Hg1201 (labeled HgUD71, $T_c = 71$~K) revealed a gapped Y-shaped spectrum both in the pseudogap (PG) and SC states, and no evidence for a resonance. The unusual response in the SC state was attributed to strong competing PG order~\cite{chan14}. Since then, charge-density-wave (CDW) order in Hg1201 was found to be particularly pronounced at this doping level~\cite{hinton16,tabis16}.

Here we study a Hg1201 sample closer to optimal doping (HgUD88; $T_c = 88$~K), motivated by early work for optimally-doped Hg1201 that yielded initial evidence for a resonance~\cite{yu10}. 
First, we confirm the observation for HgUD71~\cite{chan14} that the  response is enhanced below $T^*$, and has a gapped, Y-shaped spectrum in the PG state.
Whereas the large gap (about 40 meV) is unchanged in the SC state, 
the response of HgUD88 changes to a distinct hourglass topology and features a resonance-like enhancement at $\omega_r\sim$ 59 meV. 
This is reminiscent of the phenomenology established for the bilayer cuprates~\cite{bourges00,fauque07}.  
The characteristic resonance energy and spectral weight scale with the particle-hole Stoner continuum threshold energy in a manner consistent with results for other cuprates, and with expectations for a spin-exciton resulting from an itinerant spin formalism. 

The sample, prepared following previously described procedures~\cite{zhao06,barisic08,chan14}, consists of approximately $30$ coaligned single crystals with a total mass of $2.8$~g with full-width-at-half-maximum (FWHM) mosaic of 1.5$^{\circ}$. 
Similar to ref. \cite{chan14}, the value $T_c = 88$~K signifies the transition midpoint obtained by averaging uniform magnetic susceptibility data for the diamagnetic signal of the individual crystals.
Measurements were performed on the ARCS time-of-flight (TOF) spectrometer at Oak Ridge National Laboratory~\cite{abernathy12}, with the sample's crystalline $c$-axis aligned along the incident beam, and incident neutron energies $E_i = 100$~meV (at $5$~K, $100$~K and 250 K) and $130$~meV ($5$~K). The dynamic magnetic susceptibility, $\chi^{\prime\prime}({\bf q},\omega)$, was determined from the scattering intensity, calibrated to a Vanadium standard, by normalizing by the anisotropic Cu$^{2+}$ form factor~\cite{shamoto93} and the Bose population factor. The temperature dependence was measured
at $({\bf q},\omega) = ({\bf q_{\rm AF}},60$ meV) at the Laboratoire L{\'e}on Brillouin, with the 2T triple-axes-spectrometer, with fixed final energy $E_f=35$~meV. We quote the scattering wave-vector ${\bf Q} =H{\bf a}^* + K{\bf b}^* + L{\bf c}^*\equiv$~($H,K,L$) in reciprocal lattice units (r.l.u.), where $a^*=b^*=1.62$ \AA$^{-1}$ and $c^*=0.66$ \AA$^{-1}$ are the room-temperature values. Constant-$\omega$ data are fit to a gaussian, $\chi^{\prime\prime}({\bf q},\omega) = \chi^{\prime\prime}_0 {\rm exp}\{-4{\rm ln}2R/(2\kappa)^2\}$, convolved with the experimental momentum resolution, where {\bf q} is the reduced two-dimensional wave vector,  $R =|[(H-1/2)^2+(K-1/2)^2]^{1/2}-\delta|^2$, $2\kappa$ the intrinsic FWHM momentum width, and $\delta$ the incommensurability away from ${\bf q}_{\rm AF}$; see ref.~\cite{chan14} for further data analysis details. 

\begin{figure}[t]
\includegraphics[width=.30\textwidth]{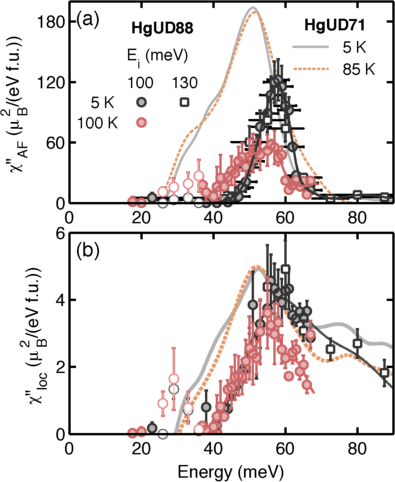}
\caption{(a) Energy dependence of magnetic susceptibility at ${\bf q}_{AF}$, $\chi^{\prime\prime}_{\rm AF}$, determined from fits to data such as those in Fig.~\ref{slice} (see text). Filled circles: $E_i = 100$~meV; the data around $\omega=30$~meV are contaminated by phonon scattering~\cite{chan14} and indicated by open circles. Open squares: $E_i = 130$~meV. Solid lines: guides to the eye. Horizontal bars represent the energy binning (only shown for $5$~K).  A large difference in $\chi^{\prime\prime}_{\rm AF}$ is observed across $T_c$. In contrast, $\chi^{\prime\prime}_{\rm AF}$ is nearly the same for HgUD71~\cite{chan14} at $5$~K and $85$~K (grey and orange lines in a) and b)). (b) Energy dependence of local susceptibility, $\chi^{\prime\prime}_{\rm loc}$. For both HgUD88 and HgUD71~\cite{chan14}, the magnetic response exhibits large gaps in the PG and SC states. \label{endep}}
\end{figure}	

Figure~\ref{slice}a-j shows $\chi^{\prime\prime}({\bf q})$ for select $\omega$ at $T = 5$, 100 and 250 K. At $5$~K, the gapped spectrum evolves with increasing energy from an incommensurate ring that disperses toward ${\bf q}_{\rm AF}$ and then outward again, thus exhibiting an archetypical hourglass dispersion (Fig.~\ref{slice}k). At $T=100$~K ($T_c+12$~K), however, the low-energy response is commensurate with ${\bf q}_{\rm AF}$ (Figs.~\ref{slice}d,i), resulting in the Y-shaped dispersion (Fig.~\ref{slice}l) that is characteristic of the PG state~\cite{chan14,bourges00, hinkov07}. Finally, at $T = 250$ K, just above $T^* \approx 220$~K~\cite{barisic13}, 
the response is considerably weaker than deep in the PG state. 

\begin{figure}[t!!!!!!!!!!!!]
\includegraphics[width=.35\textwidth]{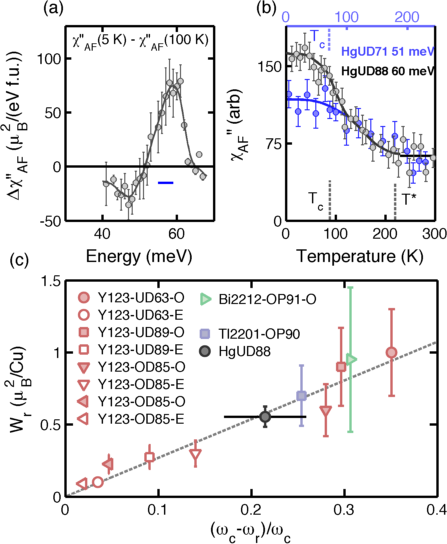}%
\caption{ (a) Change of $\chi^{\prime\prime}_{\rm AF}$ across $T_c$. Horizontal blue bar: FWHM energy resolution. The large peak at $\omega_r=59\pm1$~meV is the magnetic resonance. 
(b) Temperature dependence of $\chi^{\prime\prime}_{\rm AF}$ at $\omega\approx\omega_r$ (black) measured with a triple-axes spectrometer with FWHM energy resolution $\approx 10$ meV. $T_c$ and $T^\star$ (interpolated from planar transport measurements ~\cite{barisic13}) are indicated by the black dashed vertical lines. In contrast to HgUD88, the magnetic response of HgUD71 (blue) saturates in the SC state (from ref. \cite{chan14}); the temperature axis (top) is scaled to match $T_c$ for HgUD71 with $T_c$ for HgUD88.  
(c) $W_{\rm r}$ as a function of $\rm ( \omega_c - \omega_r) / \omega_c$ for numerous cuprates. The linear scaling and zero intercept (dashed line) are consistent with a spin-exciton description of the resonance. Y123~\cite{fong00,pailhes06} and  Bi2212~\cite{fong99} are bilayer cuprates and thus exhibit odd- and even-parity resonances, whereas single-layer Tl2201~\cite{he02} and Hg1201 (HgUD88, present work) feature only one resonance mode. Labels indicate the hole concentrations corresponding to underdoped (UD), optimally doped (OP), and overdoped (OD) regimes, followed by numbers designating $T_c$ and, if relevant, even (E) and odd (O) resonance modes.  \label{Wr}}
\end{figure}

The response at $\omega = 60$~meV (Fig.~\ref{slice}c,h), where the upward dispersion begins (Fig.~\ref{slice}g,k,l), is significantly larger at 5 K than at 100 K. 
This is reflected in a sharp peak at 5 K in the energy dependence of $\chi^{\prime\prime}_{\rm AF}$  (Fig.~\ref{endep}a).  
Detailed measurement of the temperature dependence of  $\chi^{\prime\prime}_{\rm AF}$ at 60 meV (Fig.~\ref{Wr}b) shows a distinct increase of scattering below $T^\star$, consistent with the result for HgUD71~\cite{chan14}. However, contrary to HgUD71 (Fig. 2a, 3b), this is followed by a further increase below $T_c$. 
We identify this feature below $T_c$ in HgUD88 as the magnetic resonance~\cite{mignod91,fong99,yu10}.

The resonance shows a distinct enhancement in magnetic scattering below $T_c$ in optimally- and over-doped cuprates~\cite{pailhes04b,dai01,yu10}.  
However, it is harder to discern in underdoped samples, which exhibit significant magnetic scattering in the normal state~\cite{dai01,hinkov07,chan14}, because the instrumental energy resolution is large compared to the resonance width. For HgUD88, where $\omega_r$ and hence the energy resolution of the triple-axis spectrometer are particularly large, the temperature dependence is considerably smoothed (Fig. \ref{Wr}b).

The resonance is better revealed as a peak in $\Delta\chi^{\prime\prime}_{\rm AF} =  \chi^{\prime\prime}_{\rm AF}(5{\rm~K}) - \chi^{\prime\prime}_{\rm AF}(100{\rm~K})$ (Fig.~\ref{Wr}a), centered at $\omega_r = 59 (1)$~meV, with a width that is not much larger than the experimental resolution of the TOF spectrometer (about 5 meV FWHM). The ratio $\omega_r /(k_{\rm B} T_c) = 7.9$ is the largest value reported for the cuprates~\cite{bourges05,yu09}. Using $\Delta_{SC}\approx 42(2)$ meV~\cite{vishik14,liERS13} for the SC gap amplitude, the ratio $\omega_r /\Delta_{SC}  \approx 0.70(3)$ is consistent with the value 0.64(4) established for unconventional superconductors \cite{yu09}. 

The present result for HgUD88 bears a striking resemblance to observations for bilayer Y123~\cite{bourges00,hinkov07}. The hourglass dispersion, particularly the dispersive low-energy branch, is present only below $T_c$, and thus a characteristic of the SC state. Above $T_c$, both the resonance  and its downward dispersive  branch disappear, yielding a Y-shaped spectrum~\cite{note1}. 
However, for HgUD88 the neck of the hourglass at 5~K is somewhat extended compared to other cuprates (Fig. 1f).  
Furthermore, the upper dispersion branch extends to slightly lower energy at 100~K than at 5~K. 
These subtle features, established in the TOF experiment, in combination with the coarse triple-axis energy resolution used to measure the temperature dependence, might further obscure a distinct enhancement of $\chi^{\prime\prime}_{\rm AF}(\omega_r)$ at $T_c$ (Fig. 3b). 

The energy-integrated spectral weight of the resonance peak at ${\bf q}_{AF}$  is defined as $W_r = \int d\omega  \Delta\chi^{\prime\prime}_{\rm AF}$. 
For HgUD88, we find $W_r=0.54(7)~\mu_{\rm B}^2$/Cu upon integrating from 51 to  64~meV. 
$W_r$ can be related to $\omega_r$. Within the itinerant picture, the interacting spin susceptibility is computed using the random phase approximation. 
In the SC state, the resonance at  ${\bf q}_{\rm AF}$ is part of a spin exciton, i.e., a spin-triplet collective mode bound below the threshold of the Stoner continuum, $w_c$~\cite{millis96,eschrig06}. 
The weight of the resonance is linearly related to the reduced binding energy, $(\omega_c - \omega_r) / \omega_c$, by $ W_r \simeq (g \mu_B)^2 2 \pi (V^2 \beta)^{-1} (\omega_c - \omega_r) / \omega_c$, where $V$ is the  planar interaction that enhances the bare susceptibility and $g=2$ is the Land\'e factor. 
The quantities $\omega_c$ and $\beta$ are related to the hot-spots (hs), defined as Fermi-surface points connected by {\bf q}$_{AF}$:
$\beta=4/(\pi \nu_{hs} \nu_{hs+Q_{AF}} \sin (\Theta_{hs}))$, where $\nu_{hs}$ and $\nu_{hs+Q_{AF}}$ are the Fermi velocities at the hot-spots and $\Theta_{hs}$ is the angle between their directions; $\rm \omega_c$ at the hot-spots is estimated as $1.8 \Delta_{SC}$~\cite{pailhes06}, where  $\Delta_{SC}\approx 42(2)$ meV~\cite{vishik14,liERS13}. 
As shown in Fig.~\ref{Wr}c, upon combining our result for HgUD88 with those for Y123~\cite{pailhes06}, Bi$_2$Sr$_2$CaCu$_2$O$_{8+\delta}$ (Bi2212)~\cite{fong99} and Tl2201~\cite{he02}, we find remarkably good linear scaling with zero intercept between $W_r$ and the reduced binding energy.
The common scaling factor implies universal band-structure and interaction parameters, within the experimental error, for different cuprate families and hole concentrations.

Alternatively, the resonance has been attributed to a redistribution of spectral weight of local spin fluctuations from energies below to energies above a spin gap that appears in the SC state~\cite{stock04,tranquada07}. The gap in HgUD88 is apparent from the lack of low-energy magnetic scattering (Figs.~\ref{slice}e,j). To better determine the gap size, we examine the local susceptibility, $\chi^{\prime\prime}_{\rm loc}(\omega) = \int \chi^{\prime\prime}({\bf Q},\omega) d^2{\bf q}/\int d^2{\bf q}$ (integration over the AF Brillouin zone). As seen from Fig. \ref{endep}b, HgUD88 features a particularly large  gap of about 40 meV in both the PG and SC states. 
With increasing temperature, the strength of magnetic excitations decreases, yet the gap does not close. Consistent with the result for HgUD71 \cite{chan14}, the gap thus is a property of the PG and not the SC state~\cite{dai01,stock04}. We thus cannot attribute the resonance to a spectral weight redistribution due to the opening of a gap. 
Although prior neutron scattering work yielded evidence for a ``spin-pseudogap"~\cite{mignod91,lee03,bourges05,chan14}, the present result constitutes the clearest and largest manifestation of such a gap. 

The spin-exciton scenario can semi-quantitatively account for (i) the magnitude of the resonance and (ii) its connection to a downward dispersing mode in the SC state of HgUD88. However, it fails to explain the absence of both features
in HgUD71~\cite{chan14}. It is interesting to compare the two-particle spectra in the charge and spin sectors, probed by electronic Raman scattering (ERS) and neutron scattering, respectively. In ERS, the hallmark of the SC state is the  pair-breaking peak in the $\rm B_{1g}$ channel, which probes the antinodal regions of the Fermi surface that are approximately spanned by  {\bf q}$_{AF}$. The magnitude of this peak decreases with decreasing doping. Thus, while the pair-breaking peak is sizable in HgUD88, it is much weaker in HgUD71, and disappears at lower doping~\cite{letacon06,liERS12,liERS13}. 
This phenomenon could be ascribed to the vanishing of coherent
Bogoliubov quasiparticles, because at lower doping an increasing portion of
the Fermi surface is dominated by the PG.
Furthermore, CDW order is particularly prominent in underdoped
Hg1201 with $T_c \approx 70$~K~\cite{hinton16,tabis16}, which contributes 
to the destruction of quasiparticle coherence on portions of the Fermi
surface connected by the CDW wavevector.

Our results establish that excitations across the Fermi-surface in the presence of either SC and/or PG order should be considered in accounting for the magnetic spectrum in both double- and single-layer cuprates. Recent transport measurements indicate Fermi-liquid behavior in the PG state~\cite{chan14b,mirzaei13,barisic13}, which adds further support for the need to pursue such formulations. However, the spin-exiton scenario can in principle only generate a single pole (below the Stoner continuum) at each ${\bf Q}$, 
and thus this scenario cannot account for both the downward and upward dispersive branches. In addition to the pair-breaking peak in the ERS $\rm B_{1g}$ channel, a two-magnon peak is observed~\cite{liERS13}, which indicates the persistence of short-range local-moment AF correlations, likely associated with the upward dispersive part of the spectrum. A theoretical approach that incorporates both itinerant and local spins, such as in ref.~\cite{eremin12}, might thus be necessary. Understanding the Y-shaped spectrum and the large spin gap will likely require the consideration of the relationship between the magnetic degrees of freedom and the experimentally-detected broken symmetry states~\cite{kaminski02, fauque06, li08, xia08, ghiringhelli12, tabis16} in the PG state.

We have established a phenomenology of magnetic excitations in the PG and SC states of the cuprates that is common to single- and double-layer compounds, namely a Y-shaped PG spectrum that evolves into an X-shaped (hourglass) response accompanied by a resonance in the SC state. 
In the La-based cuprates, the magnetic spectrum does not undergo a sudden X to Y transition at $T_c$. However, the low-energy incommensurability was found to decrease slowly with increasing temperature, e.g., from $\delta\approx 0.12$ at $8$~K to to $0.08$ at $200$~K in La$_{1.875}$Ba$_{0.125}$CuO$_4$~\cite{fujita04}. It is tempting to attribute this to pre-formed SC pairs, which have been argued to appear at high temperatures in the La-based cuprates~\cite{xu00}. However, the lack of a ${\bf q}_{\rm AF}$ resonance, the prominence of static stripe correlations~\cite{tranquada07}, and recent experiments indicating a narrow SC fluctuation range above $T_c$~\cite{cyr09,grbic09,bilbro11}, indicate the proximity of a stripe instability in this particular cuprate family as the dominating factor determining their low-energy magnetic spectrum.

\begin{acknowledgments}
We thank Andrey Chubukov, Yuan Li and Guichuan Yu for comments on the manuscript. 
This work was funded by the Department of Energy through the University of Minnesota Center for Quantum Materials, under DE-FG02-06ER46275 and DE-SC-0006858, and through Award No. LANLF100. LLB is supported by UNESCOS (contract ANR-14-CE05-0007)  and  NirvAna (contract ANR-14-OHRI-0010) of the ANR. ORNL's SNS is sponsored by the Scientific User Facilities Division, Office of Basic Energy Sciences, US Department of Energy. 
\end{acknowledgments}

\bibliography{UD88paper}

\end{document}